\documentclass[11pt,oneside]{book}  
 
\usepackage{amsmath,amssymb,revsymb,graphicx,dcolumn}

\begin{document}
\title{From Analogue Models to Gravitating Vacuum }
\author{
G.E.~Volovik 
\\
Low Temperature Laboratory, Aalto University, Finland\\
L.D. Landau Institute for Theoretical Physics, Moscow, Russia}

\date{\today}


\newcommand{\beq}{\begin{equation}}
\newcommand{\eeq}{\end{equation}}
\newcommand{\beqa}{\begin{eqnarray}}
\newcommand{\eeqa}{\end{eqnarray}}
\newcommand{\bsubeqs}{\begin{subequations}}
\newcommand{\esubeqs}{\end{subequations}}
\newcommand{\dd}{\mathrm{d}}                    
\newcommand{\half}{{\textstyle \frac{1}{2}}}    



\maketitle
\tableofcontents

\vspace{5mm} 

 We discuss phenomenology of quantum vacuum. Phenomenology of macroscopic systems has three sources: thermodynamics,
topology and symmetry. Momentum space topology determines the universality classes of fermionic vacua. The  vacuum in its massless state belongs to the Fermi-point universality class, which has topologically protected fermionic quasiparticles. At low energy they behave as relativistic massless Weyl fermions. Gauge fields and gravity emerge together with Weyl fermions at low energy.  Thermodynamics of the self-sustained vacuum allows us to treat the problems
related to the vacuum energy: the cosmological constant problems. The 
natural value of the energy density of the equilibrium  the self-sustained vacuum is zero.  Cosmology is the process of relaxation of vacuum towards the equilibrium state. The present value of the cosmological constant is very small compared to the Planck scale, because the present Universe is very old and thus is close to equilibrium.

\section{Introduction. Phenomenology of quantum vacuum}
\subsection{Vacuum as macroscopic many-body system}

The practical use of Analogue Gravity started 30 years ago with the remark  "that the same arguments which lead to black-hole evaporation also predict that a thermal spectrum of sound waves should be given out from the sonic horizon in transsonic fluid flow" \cite{Unruh1981}. In this paper by Unruh, the space-time geometry of the gravitational field has been modeled by the effective flow of a liquid, which plays the role of aether. The other types of analogue aether are represented by elastic media 
\cite{Bilby1956,Kroener1960,Dzyaloshinskii1980,KleinertZaanen2004,Vozmediano2010,Zaanen2010};  topological matter  with Weyl fermions \cite{Froggatt1991,Volovik2003,Horava2005};  fermionic matter experiencing the four dimensional quantum Hall effect \cite{Shou-ChengZhang2004}; etc. 

Now the aether is becoming far more than the analogy.  The aether of the 21-st century is the quantum vacuum, which is a new form of matter. This is the real substance, which however has a very peculiar properties  strikingly different from the other forms of matter (solids, liquids, gases,  plasmas, Bose condensates, radiation, etc.) and from all the old aethers. The new aether is Lorentz invariant with great accuracy
(except probably of the neutrino sector of the quantum vacuum \cite{OPERA2011}, where in principle the  Lorentz invariance can be spontaneously broken 
\cite{KlinkhamerVolovik2005b,KlinkhamerVolovik2011b}). It has equation of state $p=-\epsilon$; as follows from the  cosmological observations its energy density is about $10^{-29}$g/cm$^3$ (i.e. the quantum aether by 29 orders magnitude lighter than water); and  it  is actually anti-gravitating. 

Quantum vacuum can be viewed as a macroscopic many-body system. 
 Characteristic energy scale in our vacuum
(analog of atomic scale in quantum liquids) is Planck energy $E_{\rm P}=(\hbar c^5/G)^{1/2} \sim 10^{19}$ GeV $\sim 10^{32}$K.
Our present Universe has extremely low energies and  temperatures compared to the Planck scale:
even the highest energy in the nowadays  accelerators  is extremely small compared to Planck energy: $E_{\rm max} \sim  10$ TeV $\sim 10^{17}$K$\sim 10^{-15}E_{\rm P}$. 
The temperature of cosmic background radiation is much smaller $T_{CMBR} \sim  1$ K$\sim 10^{-32}E_{\rm P}$.

Cosmology belongs to ultra-low frequency physics. Expansion of Universe is extremely slow: the Hubble parameter compared to the characteristic Planck frequency $\omega_P=(c^5/G\hbar)^{1/2} $ is  $H  \sim 10^{-60}\omega_P$. This also means that at the moment our Universe is extremely close to equilibrium. This is natural for any many-body system: if there is no energy flux from environment the energy will be radiated away and the system will be approaching the equilibrium state with vanishing temperature and motion.

According to Landau, though the macroscopic many-body system can be very complicated,  at low energy, temperature and frequency its description is highly simplified. Its behavior can be described in a fully phenomenological way, using the symmetry and thermodynamic consideration. Later it became clear that another factor also governs the low energy properties of a macroscopic system --  topology. 
The quantum vacuum is probably a very complicated system. However, using these three sources --  thermodynamics, symmetry and topology -- one may try to construct  the phenomenological theory of the quantum vacuum near its equilibrium state.

\subsection{3 sources of phenomenology: thermodynamics, symmetry and topology}

Following Landau, at low energy $E\ll E_{\rm P}$ the macroscopic quantum system  (our Universe is an example)   contains two main components: vacuum (the ground state) and matter (fermionic and bosonic quasiparticles above the ground state). The physical laws which govern the matter component are more or less clear to us, because we are able to make experiments in the low-energy region and construct the theory. The quantum vacuum  occupies the Planckian and  trans-Planckian energy scales and it is governed by the microscopic (trans-Planckian) physics which is  still unknown.
However,  using our experience with a similar condensed matter systems we can expect  that the quantum vacuum component should also obey the  thermodynamic laws, which emerge in any macroscopically large system, relativistic or non-relativisti. This approach allows us to treat 
the   cosmological constant problems. 

Cosmological constant was introduced by Einstein \cite{Einstein1917}, and was interpreted as the energy density of the quantum vacuum \cite{Bronstein1933, Zeldovich1967}.  Astronomical observations \cite{Riess-etal1998,Perlmutter-etal1998}
confirmed the existence of cosmological constant which   value corresponds to the energy density of order $\Lambda_{\rm obs}\sim E_{\rm obs}^4$ with the characteristic  energy scale   $E_{\rm obs}\sim 10^{-3}\;\text{eV}$.
However, naive and intuitive theoretical estimation of the vacuum energy density as the zero-point energy of quantum fields  suggests that vacuum energy must have the Planck energy scale: $\sim E_{\rm P}^4 \sim 10^{120}E_{\rm obs}$. The huge disagreement between the naive expectations and observations is naturally resolved using the thermodynamics of quantum vacuum discussed in this review. We shall see that the intuitive estimation for the vacuum energy density as $\sim E_{\rm P}^4$ is  not completely crazy, but this is valid for the Universe when it is very far from equilibrium. In the fully equilibrium vacuum the relevant vacuum energy, which enters Einstein equations as cosmological constant, is zero.

The second element of the Landau phenomenological approach to macroscopic systems is symmetry.
It is in the basis of the modern theory of particle physics -- the Standard Model, and its extension to higher energy -- the Grand Unification  (GUT).  The vacuum of Standard Model and GUT  obeys the fundamental symmetries which become spontaneously broken at low energy, and are restored when the Planck energy scale is approached from below.  In the GUT scheme, general relativity is assumed to be as fundamental as quantum mechanics.

This approach contains another huge disagreement between the naive expectations and observations. It concerns masses of elementary particles. The naive and intuitive estimation tells us that these masses should be on the order of Planck energy scale:  $M_{\rm theor} \sim E_{\rm P}$, while the  masses of observed particles are many orders of magnitude smaller being below the electroweak energy scale $M_{\rm obs}<E_{\rm ew}\sim 1$  TeV $\sim 10^{-16}E_{\rm P}$.  This  is called the hierarchy problem.
There should be a general principle, which could resolve this paradox. This is the principle of emergent physics based on the topology in momentum space, which demonstrates that our  intuitive estimation of fermion masses of order $E_{\rm P}$ is not completely crazy, but it can be valid only for such vacua where the massless fermions are not protected by topology. 

\subsection{Vacuum as topological medium}

Topology operates in particular with integer numbers -- topological charges -- which do not change under small deformation of the system. The conservation of these topological charges protects the Fermi surface and another object in momentum space -- the Fermi point -- from destruction.  They survive when the interaction between the fermions is introduced and modified. When the momentum of a particle approaches  the Fermi surface or  the Fermi point its energy necessarily vanishes. Thus the topology is the main reason why there are gapless   quasiparticles in topological condensed matter  and (nearly) massless  elementary particles in our Universe. 

Topology provides the complementary anti-GUT approach in which the `fundamental' symmetry and `fundamental'  fields of GUT  gradually emerge together with `fundamental' physical laws when the Planck energy scale is approached from above \cite{Froggatt1991,Volovik2003,Horava2005}. The emergence of the `fundamental'  laws of physics is provided by the general property of topology -- robustness to details of the microscopic trans-Planckian physics. As a result, the physical laws which emerge at low energy  together with the matter itself are generic. They do not depend much on the details of the trans-Planckian subsystem, being  determined by the universality  class,  which the vacuum belongs to.  Well below the Planck scale,  the GUT scenario  intervenes: the effective symmetries which emerged due to the topology reasons exhibit spontaneous breaking at low energy. This is accompanied by formation of  composite objects, Higgs bosons, and gives tiny Dirac masses to quarks and leptons.  

In the anti-GUT scheme, fermions are primary objects. Approaching the Planck energy scale from above, they are transformed to the Standard Model chiral fermions  and  give rise to the secondary objects: gauge fields $A_\mu$ and tetrad field $e^\mu_a$. The effective metric emerges as the composite object of tetrad field,  $g^{\mu\nu}=\eta^{ab}e^\mu_a e^\nu_b$.   In this  approach, general relativity is the effective theory describing the dynamics of the effective metric experienced by the effective low-energy fermionic and bosonic fields. It is a side product of quantum field theory (or actually of the many-body quantum mechanics) in the vacuum with Fermi point. The emergence of the tetrad field before the metric field suggests that the effective theory for gravitational field must be of the Einstein-Cartan-Sciama-Kibble type \cite{Nieh2007}, which incorporates the torsion field, rather than the original Einstein theory
(see also Refs. \cite{Akama1978,Volovik1986,Wetterich2004,Diakonov2011} for the other origin of torsion field from fermions).

Vacua with topologically protected gapless (massless) fermions 
are representatives of the broader class of topological media. In condensed matter it includes
topological insulators (see review \cite{HasanKane2010}), semimetals, topological superconductors and superfluids (see review \cite{Xiao-LiangQi}), states which experience quantum Hall effect, graphene, and other topologically nontrivial gapless and gapped phases of matter. 
Topological media have many peculiar properties:  topological stability of gap nodes; topologically protected edge states including Majorana fermions; topological quantum phase transitions occurring at $T=0$; topological quantization of physical parameters including Hall and spin-Hall conductivity;  chiral anomaly; topological Chern-Simons and Wess-Zumino actions; etc. \cite{Volovik2003} 
The modern aether -- the quantum vacuum of Standard Model --  is also the topologically nontrivial medium both in its massless and massive states, and thus must experience the peculiar properties. 

In this review we mostly concentrate on the properties of quantum vacuum, which are relevant for the solution of main cosmological constant problem.  This is determined by the more universal phenomenon -- the thermodynamic behavior which emerges in any system in the limit of large number of elements. Topology in momentum space becomes important for the next steps: for dynamics of the cosmological constant.

\section{Quantum vacuum as self-sustained medium}
\label{Quantum_vacuum_self-sustained}

 \subsection{Vacuum energy and  cosmological constant} 
\label{sec:CC}

There is a  huge contribution to the vacuum energy density, which comes from the ultraviolet  (Planckian) degrees of freedom and is of order
$E^4_{\rm P} \approx \big(10^{28}\,\text{eV}\big)^4$.
The observed cosmological is  smaller by many orders of magnitude and corresponds to the
energy density of the vacuum  $\rho_{\rm vac}\sim \big(10^{-3}\,\text{eV}\big)^4$. 
  In general relativity, the cosmological constant  is arbitrary constant, and thus its smallness requires fine-tuning.
If gravitation would be a truly fundamental interaction,
it would be hard to understand why the energies stored
in the quantum vacuum would not gravitate
at all \cite{Nobbenhuis2006}.
If, however, gravitation would be only a low-energy effective interaction,
it could be that the corresponding gravitons as quasiparticles
do not feel \emph{all} microscopic degrees of freedom
(gravitons would be analogous to small-amplitude waves
at the surface of the ocean)
and that the gravitating effect of the vacuum energy density would be
effectively \emph{tuned away} and cosmological constant would be naturally small or zero  \cite{Volovik2003,Dreyer2007}. 

\subsection{Variables for Lorentz invariant vacuum} 

A particular mechanism of nullification of the relevant vacuum energy works for such vacua which have the property of a \emph{self-sustained medium} \cite{KlinkhamerVolovik2008a,KlinkhamerVolovik2008b,KlinkhamerVolovik2009b,KlinkhamerVolovik2009a,KlinkhamerVolovik2010}.
A self-sustained vacuum is a medium with a definite macroscopic
volume even in the absence of an environment. A condensed matter
example is a droplet of quantum liquid at zero temperature in empty space.
The observed near-zero value of the cosmological constant
compared to Planck-scale values 
suggests that the quantum vacuum of our universe
belongs to this class of systems.
As any medium of this kind, the equilibrium vacuum
would be homogeneous and extensive. The homogeneity assumption is
indeed supported by the observed flatness and
smoothness of our universe \cite{de Bernardis2000,Hinshaw2007,Riess2007}.
The implication is that the energy of the equilibrium quantum
vacuum would be proportional to the volume considered.

Usually, a self-sustained medium is characterized by an
\emph{extensive conserved quantity}
whose total value determines the actual volume of the
system \cite{LL1980,Perrot1998}. In condensed matter, the quantum liquid at $T=0$ is a self sustained system because of the conservation law for the particle number $N$, and its state is characterized by the particle density $n$
which acquires a non-zero value $n=n_0$ in the equilibrium ground state. As distinct from condensed matter systems, the 
quantum vacuum of our Universe obeys the  relativistic invariance with great accuracy.
The Lorentz invariance of the vacuum imposes strong constraints on the possible form this variable can take: the particle density $n$  must be zero in the Lorentz invariant vacuum,
since it represent the time component of the 4-vector. One must find the relativistic analog of the quantity $n$, which is invariant under Lorentz transformation. 
An example of a possible vacuum variable is a
symmetric tensor $q^{\mu\nu}$, which in a homogeneous vacuum 
is proportional to the metric tensor
\begin{equation}
q^{\mu\nu}=q\,g^{\mu\nu} \,.
\label{eq:2tensor}
\end{equation}
This variable satisfies the Lorentz invariance of the vacuum.
Another example is the 4-tensor $q^{\mu\nu\alpha\beta}$, which in a homogeneous vacuum 
is proportional either to the fully antisymmetric Levi--Civita tensor:
 \begin{equation}
q^{\mu\nu\alpha\beta}=q\,e^{\mu\nu\alpha\beta}\,, 
\label{eq:2tensor}
\end{equation}
or to the product of metric tensors such as:
\begin{equation}
q^{\mu\nu\alpha\beta}=q \left(g_{\alpha\mu} g_{\beta\nu} -g_{\alpha\nu} g_{\beta\mu}\right)
 \,.
\label{eq:mixed}
\end{equation}

Scalar field is also the Lorentz invariant variable, but it does not satisfy another necessary condition of the self sustained system: the vacuum variable $q$ must obey some kind of the conservation law. Below we consider some examples satisfying the two conditions: Lorentz invariance of the perfect vacuum state and the conservation law.

\subsection{Yang-Mills chiral condensate as example}

Let us first consider as an example the chiral condensate of gauge fields. It can be the
gluonic condensate in QCD  \cite{Shifman1992,Shifman1991}, or any other condensate of
Yang-Mills fields, if it is Lorentz invariant.
We assume that the Savvidy vacuum \cite{Savvidy}
is absent, i.e. the vacuum expectation value of the
 color magnetic field is zero (we shall omit  color indices):
\begin{equation}
\left<F_{\alpha\beta}\right>=0 \,,
\label{eq:VEVF}
\end{equation}
while the vacuum expectation value of the quadratic form is nonzero:
\begin{equation}
 \left<F_{\alpha\beta}  F_{\mu\nu} \right>=  \frac{q}{24}  \sqrt{-g}e_{\alpha\beta\mu\nu}  \,.
\label{eq:VEVF^2}
\end{equation}
Here $q$ is  the anomaly-driven topological condensate (see e.g. \cite{HalperinZhitnitsky1998}):
\begin{equation}
q= \left< \tilde F^{\mu\nu}F_{\mu\nu}\right>
= \frac{1}{\sqrt{-g}}e^{\alpha\beta\mu\nu} \left<F_{\alpha\beta}  F_{\mu\nu} \right> \,,
 \label{eq:q}
\end{equation}
In the homogeneous static vacuum state, the $q$-condensate
violates the $P$ and $T$ symmetries of the vacuum,
but it conserves the combined symmetry $PT$  symmetry.

\subsubsection{Cosmological term}

Let us choose the vacuum action in the form
\begin{equation}
 S_q=\int d^4x \sqrt{-g}\epsilon(q)\,,
\label{eq:VacuumAction}
\end{equation}
with $q$ given by \eqref{eq:q}.
The energy-momentum tensor of the vacuum field $q$ is obtained by variation
of the action over $g^{\mu\nu}$:
\begin{equation}
T^q_{\mu\nu}=-\frac{2}{\sqrt{-g}}\: \frac{\delta S_q}{\delta g^{\mu\nu}}=
\epsilon(q)\,  g_{\mu\nu} -
2\,  \frac{\partial\epsilon}{\partial q}\, \frac{\partial q}{\partial g^{\mu\nu}}
 \,.
\label{eq:em_tensor_pot}
\end{equation}
Using \eqref{eq:VEVF^2} and \eqref{eq:q} one obtains
\begin{equation}
 \frac{\partial q}{\partial g^{\mu\nu}}
=
\frac{1}{2} q  g_{\mu\nu}
 \,.
\label{eq:dq/dg}
\end{equation}
and thus
\begin{equation}
T_{\mu\nu}^q= g_{\mu\nu} \rho_{\rm vac}(q)~~,~~ \rho_{\rm vac}(q)=\epsilon(q)
- q  \frac{\partial\epsilon}{\partial q}   \,.
\label{eq:tilde_epsilon}
\end{equation}
In Einstein equations this energy momentum tensor plays the role
of the cosmological term: 
\begin{equation}
T_{\mu\nu}^q= \Lambda g_{\mu\nu} ~~,~~ \Lambda=\rho_{\rm vac}(q)= \epsilon(q)
- q  \frac{\partial\epsilon}{\partial q}   \,.
\label{eq:cosmological_term}
\end{equation}
It is important that the cosmological constant
is given not by the vacuum energy as is usually assumed, but by the equivalent of the grand potential in condensed matter systems -- the thermodynamic potential
$\rho_{\rm vac}=\epsilon(q)
 - \mu q$, where $\mu$ is thermodynamically conjugate to $q$ variable, $\mu=d\epsilon/dq$. 
 Below, when we consider dynamics, we shall see that this fact reflects the conservation of the variable $q$. 
 
  The crucial difference between the vacuum energy $\epsilon(q)$ and thermodynamic potential
$\rho_{\rm vac}=\epsilon(q)- \mu q$ is revealed when we consider the corresponding quantities in the ground state of quantum liquids, the energy density
 $\epsilon(n)$ and the grand potential
$\epsilon(n)- \mu n$. The first one, $\epsilon(n)$, has the value dictated by atomic physics, which is equivalent to $E_{\rm P}^4$ in quantum vacuum. On the contrary, the second one equals minus pressure, $\epsilon(n)- \mu n=-P$, according to the Gibbs-Duhem thermodynamic relation at $T=0$. Thus its value is dictated not by the microscopic physics, but by external conditions. In the absence of environment, the external pressure is zero, and the value  of $\epsilon(n)- \mu n$ in a fully equilibrium ground state of the liquid is zero. This is valid for any macroscopic system, and thus should be applicable to the self-sustained quantum vacuum, which suggests the natural solution of the main cosmological constant problem.

\subsubsection{Conservation law for $q$}

Equation for $q$ in flat space
can be obtained from Maxwell equation, which in turn is obtained by variation 
of the action over the gauge field $A_\mu$:
\begin{equation}
\nabla_\mu \left(\frac{\partial\epsilon}{\partial q}\tilde F^{\mu\nu} \right)=0
 \,.
\label{eq:Maxwell}
\end{equation}
Since  $\nabla_\mu \tilde F^{\mu\nu}=0$, equation \eqref{eq:Maxwell}
is reduced to
\begin{equation}
\nabla_\mu \left(\frac{\partial\epsilon}{\partial q}\right) =0 \,.
\label{eq:tilde_mu1}
\end{equation}
The solution of this equation is
\begin{equation}
 \frac{\partial\epsilon}{\partial q}=  \mu\,,
\label{eq:tilde_mu2}
\end{equation}
 where $\mu$ is integration constant. In thermodynamics, this $\mu$ will play   the role of the chemical potential, which is thermodynamically conjugate to $q$. This demonstrates that $q$ obeys the conservation law and thus can be the proper variable for description the self-sustained vacuum.

\subsection{4-form field as example}

Another example of the vacuum variable appropriate for the self-sustained vacuum is given by the four-form field
strength~\cite{DuffNieuwenhuizen1980,Aurilia-etal1980,Hawking1984,HenneauxTeitelboim1984,
Duff1989,DuncanJensen1989,BoussoPolchinski2000,Aurilia-etal2004,Wu2008}, which is expressed in
terms of $q$ in the following way:
\bsubeqs\label{eq:EinsteinF-all}
\beqa
F_{\alpha\beta\gamma\delta}  &\equiv&
q\,e_{\alpha\beta\gamma\delta}\, \sqrt{-\det g}
= \nabla_{[\alpha}A_{\beta\gamma\delta]}\,,
\label{eq:Fdefinition1}\\[2mm]
q^2 &=&
- \frac{1}{24}\,F_{\alpha\beta\gamma\delta}\,F^{\alpha\beta\gamma\delta}\,,
\label{eq:q2definition}
\eeqa
\esubeqs
where $e_{\alpha\beta\gamma\delta}$ the Levi--Civita tensor density;
$\nabla_\alpha$ the covariant derivative;
and the square bracket around spacetime indices complete anti-symmetrization.

Using vacuum action \eqref{eq:VacuumAction} with the above $q$
one obtains the Maxwell equation
\begin{equation}
\nabla_\alpha \left(\sqrt{-\det g} \,\;\frac{F^{\alpha\beta\gamma\delta}}{q}
           \frac{\partial\epsilon(q)}{\partial q} \right)=0\,.
\label{eq:4Maxwell}
\end{equation}
This equation reproduces the known equation~\cite{DuffNieuwenhuizen1980,Aurilia-etal1980}
for the special case $\epsilon(q)=\half\,q^2$.
Using \eqref{eq:Fdefinition1} the Maxwell equation is reduced to
\begin{equation}
\nabla_\alpha \left(
           \frac{\partial\epsilon(q)}{\partial q} \right)=0\,.
\label{eq:4Maxwell2}
\end{equation}
The first integral of \eqref{eq:4Maxwell2} with integration constant $\mu$
gives again Eq.\eqref{eq:tilde_mu2}, which reflects the conservation law for $q$.

Variation of the action over $g^{\mu\nu}$ gives again the cosmological constant \eqref{eq:cosmological_term} with $\Lambda=\rho_{\rm vac}=\epsilon(q)- \mu q$.
This demonstrates the universality of the description of the self-sustained vacuum: description of the quantum vacuum in terms of the field $q$ does not depend on the microscopic origin of this field. 

\subsection{Aaether field as example}
\label{Aaether_field}

Another example of the vacuum variable $q$ may be through
a four-vector field $u^{\mu}(x)$. This vector field could be the
four-dimensional analog of the concept of shift in the deformation
theory of crystals. (Deformation theory can be described in terms of a
metric field, with the role of  torsion and curvature fields played by
dislocations and disclinations, respectively
\cite{Bilby1956,Kroener1960,Dzyaloshinskii1980,KleinertZaanen2004,Vozmediano2010}).
A realization of $u^{\mu}$ could be also a 4--velocity field
entering the description of the structure of spacetime.
It is  the 4-velocity of  ``aether''  \cite{Jacobson2007,Gasperini1987,Jacobson2001,WillNordvedt}.

The nonzero value of the 4-vector in the vacuum violates
the  Lorentz invariance of the vacuum. To restore this invariance one may assume that
$u^{\mu}(x)$ is not an observable variable, instead the observables are its covariant derivatives $\nabla_\nu u^{\mu}\equiv u_\nu^\mu$. This means that the action does not depend on $u^{\mu}$ explicitly but only
depends on  $u_\nu^\mu$:
\begin{equation}
S= 
\int_{\mathbb{R}^4} \,d^4x\, 
\epsilon(u_\nu^\mu)\,,
\label{eq:action3}
\end{equation}
with an energy density containing even powers of
$u_\nu^\mu$:
\begin{equation}\label{eq:epsilon-u-mu-nu}
\epsilon(u_\nu^\mu)
= K
+ K_{\mu\nu}^{\alpha\beta}\,u_\alpha^\mu u_\beta^\nu
+ K_{\mu\nu\rho\sigma}^{\alpha\beta\gamma\delta}\,
u_\alpha^\mu u_\beta^\nu u_\gamma^\rho u_\delta^\sigma + \cdots \;.
\end{equation}
According to the imposed conditions, the
tensors $K_{\mu\nu}^{\alpha\beta}$ and
$K_{\mu\nu\rho\sigma}^{\alpha\beta\gamma\delta}$
depend only on $g_{\mu\nu}$ or $g^{\mu\nu}$ and the same holds for the other
$K$--like tensors in the ellipsis of \eqref{eq:epsilon-u-mu-nu}.
In particular, the tensor $K_{\mu\nu}^{\alpha\beta}$ of the quadratic term
in \eqref{eq:epsilon-u-mu-nu} has the following form in the notation of
Ref.~\cite{Jacobson2007}:
\begin{equation}
K_{\mu\nu}^{\alpha\beta}=c_1\, g^{\alpha\beta}g_{\mu\nu}  +
c_2\, \delta^{\alpha}_{\mu}  \delta^{\beta}_{\nu}  +
c_3\, \delta^{\alpha}_{\nu}  \delta^{\beta}_{\mu}  ~,
\label{eq:Kparameters}
\end{equation}
for real constants $c_n$.
Distinct from the original aether theory in Ref.~\cite{Jacobson2007},
the tensor \eqref{eq:Kparameters}
does not contain a term $c_4\, u^\alpha u^\beta g_{\mu\nu}$,
as such a term would depend explicitly on $u^{\mu}$ and
contradict the Lorentz invariance
of the quantum vacuum.

The equation of motion for $u^{\mu}$ in flat space,
\begin{equation}
\nabla_\nu  \, \frac{\partial\epsilon}{\partial u_\nu^\mu} =0\,,
\label{eq:Motion}
\end{equation}
has the Lorentz invariant solution expected for a vacuum-variable $q$--type field:
\begin{equation}
u^q_{\mu\nu}=q\,g_{\mu\nu}\,,\quad q=\text{constant}~.
\label{eq:solution3}
\end{equation}
With this solution, the energy density in the action \eqref{eq:action3} is
simply $\epsilon(q)$
in terms of contracted coefficients $K$, $K_{\mu\nu}^{\mu\nu}$, and
$K_{\mu\nu\rho\sigma}^{\mu\nu\rho\sigma}$ from \eqref{eq:epsilon-u-mu-nu}.
However, just as for previous examples,
the energy-momentum tensor of the vacuum field obtained by
variation over $g^{\mu\nu}$ and evaluated for solution \eqref{eq:solution3}
is expressed again in terms of  the thermodynamic potential:
\beqa
T^q_{\mu\nu}
&=&
\frac{2}{\sqrt{-g}}\;\frac{\delta S}{\delta g^{\mu\nu}}\,
 =              g_{\mu\nu} \left(\epsilon(q) - q\,\frac{d\epsilon(q)}{dq}\right)=\rho_{\rm vac}(q) g_{\mu\nu}\,,
\label{eq:emSolution3}
\eeqa
which corresponds to cosmological constant in Einstein's gravitational field
equations.

\section{Thermodynamics of quantum vacuum}

\subsection{Liquid-like quantum vacuum}

\begin{figure}
\centerline{\includegraphics[width=0.8\linewidth]{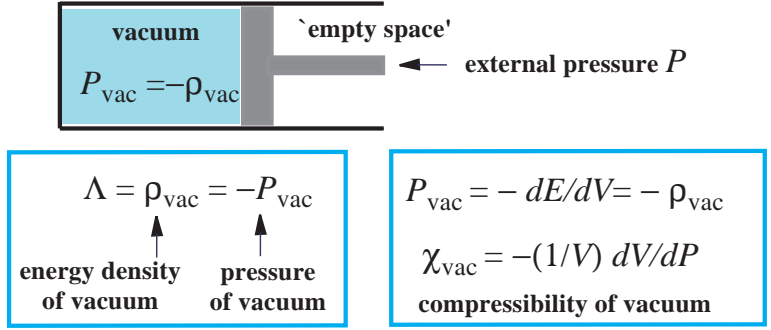}}
\caption{Vacuum as a medium obeying macroscopic thermodynamic laws. Relativistic vacuum possesses energy density, pressure and compressibility but has no momentum.  In equilibrium, the vacuum pressure $P_{\rm vac}$ equals the external pressure $P$ acting from the environment. 
The  ``gas-like'' vacuum may exist only under external pressure. The  ``liquid-like'' vacuum is self-sustained: it can be stable in the absence of external pressure.
The thermodynamic energy density of the vacuum $\rho_{\rm vac}$ which enters the vacuum equation of state $\rho_{\rm vac}= -P_{\rm vac}$ does not coincide with the microscopic vacuum energy $\epsilon$. While the natural value of  $\epsilon$ is determined by the Planck scale, $\epsilon\sim E_{\rm P}^4$,  the natural value of the macroscopic quantity $\rho_{\rm vac}$ is zero for the self-sustained vacuum which may exist in the absence of environment, i.e. at $P=0$. This may explains why the present cosmological constant $\Lambda=\rho_{\rm vac}$ is small.
}  
\label{compressibility} 
\end{figure}

The zeroth order term $K$ in \eqref{eq:epsilon-u-mu-nu} corresponds to a ``bare'' cosmological constant which can be considered as 
cosmological constant in the ``empty'' vacuum -- vacuum with $q=0$:
\begin{equation}
\Lambda_{\rm bare}=\epsilon(q=0)~.
\label{eq:Lambda_bare}
\end{equation}
 The nonzero  value $q=q_0$ in the self-sustained vacuum does \emph{not} violate Lorentz symmetry
but leads to compensation of the bare cosmological constant $\Lambda_{\rm bare}$
in the equilibrium vacuum.
This  illustrates the important difference between the 
two states of vacua. The quantum vacuum with $q=0$ 
can exist only with external pressure $P =-\Lambda_{\rm bare}$.
By analogy with condensed-matter physics,
this kind of quantum vacuum may be called ``gas-like'' (Fig. \ref{compressibility}).
The quantum vacuum with nonzero $q$
is self-sustained: it can be stable at $P=0$, provided that a stable nonzero solution of equation
$\epsilon(q) - q\, d\epsilon/dq=0$ exists.
This kind of quantum  vacuum may then be called ``liquid-like''.

The universal behavior of the self-sustained vacuum in equilibrium suggests that it obeys the same thermodynamic laws as any other 
self-sustained macroscopic system described by the conserved quantity $q$, such as
quantum liquid.
In other words, vacuum can be considered as a special quantum medium which is Lorentz invariant 
in its ground state.  This medium is characterized by  the Lorentz invariant ``charge'' density $q$ -- an analog of particle density $n$ in non-relativistic quantum liquids.

Let us consider  a large portion of such  vacuum medium under external pressure $P$ \cite{KlinkhamerVolovik2008a}. The volume $V$ of quantum vacuum is variable, but its total ``charge''
$Q(t)\equiv \int d^3r~q(\mathbf{r},t)$ must be conserved,
$\mathrm{d}Q/\mathrm{d}t=0$. 
The energy of this portion of quantum vacuum at fixed  total``charge'' $Q=q\, V$
is then given by the thermodynamic potential
\begin{equation}
W=E+P\,V=\int d^3r~\epsilon\left(Q/V\right) + P\,V \,,
\label{eq:ThermodynamicPotential}
\end{equation}
where
$\epsilon\left(q\right)$ is the energy density in terms of charge density $q$.
As the volume of the system is a free parameter,
the equilibrium state of the system is obtained by variation over the volume $V$:
\begin{equation}
\frac{d W}{dV}=0 \,.
\label{eq:Equilibrium}
\end{equation}
This gives an integrated form of the Gibbs--Duhem equation for the vacuum pressure:
\begin{equation}
P_{\rm vac}=-\epsilon(q) +q\,\frac{d\epsilon(q)}{dq}=-\rho_{\rm vac}(q)~,
\label{eq:Gibbs-Duhem}
\end{equation}
whose solution determines the equilibrium value $q=q(P)$
and the corresponding volume  $V(P,Q)=Q/q(P)$.

\subsection{Macroscopic energy of quantum vacuum} 

Since the vacuum energy density is the vacuum pressure with minus sign, equation (\ref{eq:Gibbs-Duhem}) suggests that the relevant  vacuum energy, which is revealed in thermodynamics and dynamics of the low-energy Universe, is the equivalent of the grand potential:
\begin{equation}
\rho_{\rm vac}(q)=\epsilon(q) -q\,\frac{d\epsilon(q)}{dq}~.
\label{eq:vev}
\end{equation}
This is confirmed by Eqs.~\eqref{eq:cosmological_term}  and \eqref{eq:emSolution3} for energy-momentum tensor of the 
self-sustained vacuum, which demonstrates
that it is the grand potential $\rho_{\rm vac}\left(q\right)$ rather than the energy density $\epsilon\left(q\right)$, which enters the equation of state for the vacuum and thus corresponds to the cosmological constant:
\begin{equation}
\Lambda=\rho_{\rm vac}=-P_{\rm vac}~.
\label{eq:EoS}
\end{equation}

While the energy of microscopic quantity $q$ is determined by the Planck scale,
$\epsilon(q_0) \sim E_{\rm P}^4$, the relevant vacuum energy which sources the effective gravity is determined by a macroscopic quantity -- the external pressure.
In the absence of an environment, i.e. at zero external pressure, $P=0$, one obtains that the pressure of pure and equilibrium vacuum is exactly zero:
\begin{equation}
\Lambda=-P_{\rm vac}=-P=0~.
\label{eq:Null}
\end{equation}
Equation $\rho_{\rm vac}(q)=0$ determines the equilibrium value $q_0$ of the 
equilibrium self-sustained vacuum. 
Thus from the thermodynamic arguments it follows that for any effective theory of gravity the natural value of $\Lambda$ is zero in equilibrium vacuum. 

 \begin{figure}
\centerline{\includegraphics[width=0.7\linewidth]{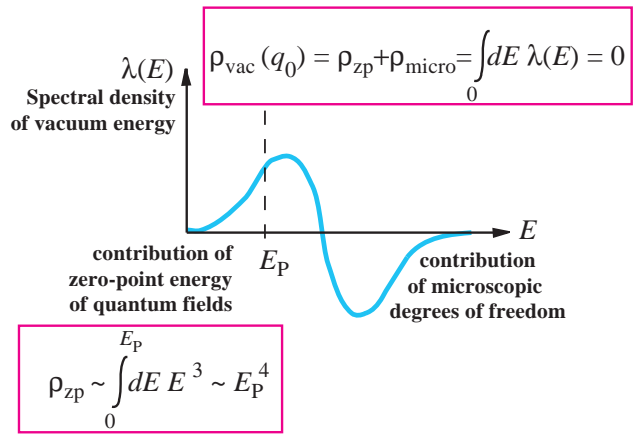}}
\caption{Contribution of different energy scales into the macroscopic energy of the self-sustained system at $T=0$. Zero point energy $\rho_{\rm zp}$ of the effective bosonic and fermionic quantum fields gives rise to the diverging contribution to the energy  of the system. In quantum vacuum it is of order of $E_{\rm P}^4$. In equilibrium this contribution is compensated without  fine-tuning by microscopic degrees of freedom of the system  (by trans-Planckian degrees of quantum vacuum correspondingly). }  
\label{spectrum} 
\end{figure}

This result does not depend on the microscopic structure of the vacuum  from which gravity emerges, and is actually the final result of the renormalization dictated by macroscopic physics. Note that we have two types of cancellation. The Planck scale quantities $\epsilon(q)$ and $q d\epsilon/dq$ cancel each other in equilibrium. The same occurs with the Planck scale contribution of zero-point energy of the bosonic and fermionic matter fields to the vacuum energy $\rho_{\rm vac}$, it is naturally compensated by microscopic degrees of freedom of the self sustained  quantum vacuum. 
The vacuum variable $q$ is adjusted automatically to nullify the relevant vacuum energy,  $\rho_{\rm vac}(q_0)= \rho_{\rm zp} +\rho_{\rm micro}=0$. The actual spectrum of the vacuum energy density (meaning the
different contributions to $\rho_{\rm vac}$ from different energy scales)
is not important for the cancellation mechanism, because it is dictated by thermodynamics. The particular example of the spectrum of the vacuum energy density is shown in Fig. \ref{spectrum}, where the positive energy of the quantum vacuum, which comes from the
zero-point energy of bosonic fields, is compensated by negative contribution from trans-Planckian degrees of freedom \cite{VolovikSpectrum}. 

Using the quantum-liquid counterpart of the self-sustained quantum vacuum as example, one may predict the behavior of the vacuum after  cosmological phase transition, when $\Lambda$ is kicked from its zero value. The vacuum will readjust itself to a new equilibrium state with new $q_0$ so that $\Lambda$ will again approach its equilibrium zero value  \cite{KlinkhamerVolovik2008a}.
This process depends on details of dynamics of the vacuum variable $q$, and later on we shall consider some examples of dynamical relaxation of $\Lambda$.

 \subsection{Compressibility of the vacuum}

Using the standard definition of the inverse of
the isothermal compressibility, $\chi^{-1} \equiv -V\,dP/dV$ (Fig. \ref{compressibility}),  one obtains the compressibility of the vacuum by varying Eq.(\ref{eq:Gibbs-Duhem}) at fixed $Q=qV$ 
\cite{KlinkhamerVolovik2008a}:
\begin{equation}
\chi_\text{vac}^{-1} \equiv -V\frac{dP_{\rm vac}}{dV}=\left[q^2\;\frac{d^2\epsilon(q)}{dq^2}\,\right]_{q=q_0}
> 0~.
\label{eq:Stability}
\end{equation}
A positive value of the vacuum compressibility
is a necessary condition for the stability of the vacuum. It is, in fact, the stability of
the vacuum, which is at the origin of the nullification of the cosmological
constant 
in the absence of an external environment.

From the low-energy point of view, the compressibility of the vacuum $\chi_\text{vac}$ is as fundamental physical constant as the Newton constant $G_N=G(q_0)$. It enters equations describing  the response of the quantum vacuum  to different perturbations $\chi_\text{vac}$.  While the natural value of the macroscopic quantity $P_{\rm vac}$ (and $\rho_{\rm vac}$) is zero, the natural values of  the parameters $G(q_0)$ and $\chi_\text{vac}(q_0)$ are determined by the Planck physics and are expected to be of order $1/E^{2}_{\rm P}$ and $1/E^{4}_{\rm P}$ correspondingly.

 \subsection{Thermal fluctuations of $\Lambda$ and the volume of Universe}

The compressibility of the vacuum $\chi_\text{vac}$, though not measurable at the moment, can be used for estimation of  the lower limit for the volume $V$ of the Universe. This estimation follows from the upper limit for thermal fluctuations of cosmological constant \cite{Volovik2004}. The mean square of thermal  fluctuations of $\Lambda$ equals the mean square of thermal  fluctuations of the vacuum pressure, which in turn is determined by thermodynamic equation \cite{LL1980}:
\begin{equation}
\left <\left(\Delta\Lambda\right)^2\right>=\left <\left(\Delta P\right)^2\right>=\frac{ T}{V\chi_{\rm vac}}~.
\label{Fluctuations}
\end{equation}
 Typical fluctuations of the cosmological constant $\Lambda$ should not exceed the observed value:  $\left <\left(\Delta\Lambda\right)^2\right>< \Lambda_{\rm obs}^2$. Let us assume, for example, that the temperature of the Universe is determined by the temperature $T_{\rm CMB}$ of the cosmic microwave background radiation. Then, using our estimate for vacuum compressibility $\chi_{\rm vac}^{-1}\sim E^4_{\rm P}$, one obtains that the volume $V$ of our Universe highly exceeds the Hubble volume $V_H=R_H^3$ --  the volume of visible Universe inside the present cosmological horizon:
\begin{equation}
V> \frac{T_{\rm CMB}} { \chi_{\rm vac} \Lambda_{\rm obs}^2}\sim 10^{28}V_H~.
\label{Volume}
\end{equation}
This demonstrates that the real volume of the Universe is certainly not limited by the present cosmological horizon. 
  
\section{Dynamics of quantum vacuum}

\subsection{Action}\label{sec:Action}

In   section \ref{Quantum_vacuum_self-sustained} a special quantity,
the  vacuum ``charge''  $q$, was introduced to describe the statics and thermodynamics of the self-sustained quantum vacuum. Now we can extend this approach to the dynamics of the vacuum charge.
We expect to find some universal features of the vacuum dynamics, using several realizations 
of this vacuum variable. We start with the 4-form field
strength~\cite{DuffNieuwenhuizen1980,Aurilia-etal1980,Hawking1984,HenneauxTeitelboim1984,
Duff1989,DuncanJensen1989,BoussoPolchinski2000,Aurilia-etal2004,Wu2008} expressed in
terms of $q$. 
The low-energy effective action takes the following general form: 
\bsubeqs\label{eq:EinsteinF-all} \beqa S=- \int_{\mathbb{R}^4}
\,d^4x\, \sqrt{|g|}\,\left(\frac{R}{16\pi G(q)} +\epsilon(q)
+\mathcal{L}^\text{M}(q,\psi)\right) \,,  
\label{eq:actionF}\\[2mm]
q^2 \equiv- \frac{1}{24}\,
F_{\kappa\lambda\mu\nu}\,F^{\kappa\lambda\mu\nu}\,,\quad
F_{\kappa\lambda\mu\nu}\equiv
\nabla_{[\kappa}\!\!A_{\lambda\mu\nu]}\,,  
\label{eq:Fdefinition}\\[2mm]
F_{\kappa\lambda\mu\nu}=q\sqrt{|g|} \,e_{\kappa\lambda\mu\nu}\,,\quad
F^{\kappa\lambda\mu\nu}=q \,e^{\kappa\lambda\mu\nu}/\sqrt{|g|}\,. \quad
\label{eq:Fdefinition2} 
\eeqa \esubeqs
where $R$ denotes the Ricci curvature scalar; and $\mathcal{L}^\text{M}$ is matter action.
The vacuum energy density $\epsilon$ in \eqref{eq:actionF}
depends on the vacuum variable $q$ which in turn is expressed via the 3-form field 
$A_{\lambda\mu\nu}$ and metric field $g_{\mu\nu}$ in \eqref{eq:Fdefinition}.
The field $\psi$ combines all the matter fields of the Standard Model. All possible constant terms in matter action (which includes the zero-point energies
from the Standard Model fields) are absorbed in the vacuum energy $\epsilon(q)$.

Since $q$ describes the state of the vacuum, the parameters of the effective action -- the Newton constant $G$ and parameters which enter the matter action -- must depend on $q$.  This
dependence results in particular in the interaction between the matter fields and the vacuum. There are
different sources of this interaction. For example, in the gauge field sector of Standard Model, the running
coupling contains the ultraviolet cut-off and thus depends on $q$:
\begin{equation}
\mathcal{L}^\text{{\bf G},q}= \gamma(q)F^{\mu\nu}F_{\mu\nu}
\,, \label{RunningCupling}
\end{equation}
where $F_{\mu\nu}$ is the field strength of the particular gauge field (we omitted the color indices).
In the fermionic sector, $q$ should enter parameters
of the Yukawa interaction and fermion masses.

\subsection{Vacuum dynamics}

The variation of the action \eqref{eq:actionF} over the three-form
gauge field $A$ gives the generalized Maxwell equations for $F$-field,
\begin{equation}
\nabla_\nu \left(\sqrt{|g|} \;\frac{F^{\kappa\lambda\mu\nu}}{q} \left(
\frac{d\epsilon(q)}{d q}+\frac{R}{16\pi} \frac{dG^{-1}(q)}{d q}
+ \frac{d \mathcal{L}^\text{M}(q)}{d q}
\right)\right)=0\,.
\label{eq:Maxwell}
\end{equation}
Using \eqref{eq:Fdefinition2} for $F^{\kappa\lambda\mu\nu}$,
we find that the solutions of Maxwell equations \eqref{eq:Maxwell} 
are still determined by the integration constant $\mu$
\begin{equation}
\frac{d\epsilon(q)}{d q}+\frac{R}{16\pi} \frac{dG^{-1}(q)}{d q}
+ \frac{d \mathcal{L}^\text{M}(q)}{d q}
=\mu \,.
\label{eq:Maxwell2}
\end{equation}

\subsection{Generalized Einstein equations}

The variation over the metric $g^{\mu\nu}$ gives the
generalized Einstein equations,
\begin{eqnarray}
&&
\frac{1}{8\pi G(q)}
\left( R_{\mu\nu}-\frac{1}{2}\,R\,g_{\mu\nu}\right)
+\frac{1}{16\pi}\, q\,\frac{d G^{-1}(q)}{d q}\, {R}\,g_{\mu\nu}
\nonumber\\[2mm]
&&+ \frac{1}{8\pi} \Big( \nabla_\mu\nabla_\nu\, G^{-1}(q) - g_{\mu\nu}\,
\Box\, G^{-1}(q)\Big) -\left( \epsilon(q) -q\,\frac{d\epsilon(q)}{d q}\right)g_{\mu\nu}
\nonumber\\[2mm]
&&+  q\frac{\partial \mathcal{L}^\text{M}}{\partial q} g_{\mu\nu}
 +T^\text{M}_{\mu\nu} =0\,, \label{eq:EinsteinEquationF}
\end{eqnarray}
where $\Box$ is the invariant d'Alembertian;
and
$T^\text{M}_{\mu\nu}$ is the  energy-momentum tensor of the matter
fields, obtained by variation over $g^{\mu\nu}$ at constant $q$,
i.e. without variation over
 $g^{\mu\nu}$, which enters $q$.

Eliminating $dG^{-1}/dq$ and $\partial \mathcal{L}^\text{M}/\partial q$ from \eqref{eq:EinsteinEquationF} by use of
\eqref{eq:Maxwell2}, the generalized Einstein equations become
\begin{equation}
\frac{1}{8\pi G(q)}\Big( R_{\mu\nu}-\half\,R\,g_{\mu\nu} \Big) 
+ \frac{1}{8\pi}
\Big( \nabla_\mu\nabla_\nu\, G^{-1}(q) - g_{\mu\nu}\, \Box\, G^{-1}(q)\Big)
- \rho_{\rm vac} g_{\mu\nu}+T^\text{M}_{\mu\nu} =0\,,
\label{eq:EinsteinEquationF2}
\end{equation}
where
\begin{equation}
\rho_{\rm vac}=\epsilon(q)-\mu\, q \,.
\label{eq:Lambda}
\end{equation}
For the special case when the dependence of the Newton constant and matter action on $q$ is ignored, \eqref{eq:EinsteinEquationF2}
reduces to the standard Einstein equation of general relativity with the constant cosmological constant $\Lambda= \rho_{\rm vac}$.

\subsection{Minkowski-type solution and Weinberg problem}

Among different solutions of equations \eqref{eq:Maxwell} and \eqref{eq:EinsteinEquationF}
 there is the solution corresponding to perfect equilibrium Minkowski vacuum without matter.
It is characterized by
the constant in space and time values $q=q_0$ and $\mu=\mu_0$ obeying
the following two conditions:
\bsubeqs\label{eq:equil-eqs}
\beqa
\Bigg[\, \frac{\dd \epsilon(q)}{\dd q} - \mu\, \Bigg]_{\mu=\mu_0\,,\,q=q_0} &=&0\,,
\label{eq:equil-eqs-mu}
\\[2mm]
\Big[\, \epsilon(q)    -  \mu\, q\,\Big]_{\mu=\mu_0\,,\,q=q_0} &=&0\,.
\label{eq:equil-eqs-GDcondition}
\eeqa \esubeqs
The two conditions \eqref{eq:equil-eqs-mu}--\eqref{eq:equil-eqs-GDcondition}
can be combined into a  \emph{single} equilibrium condition for $q_0$:
\beq
\Lambda_0 \equiv
\Bigg[\, \epsilon(q)  -  q\,\frac{\dd \epsilon(q)}{\dd q}\, \Bigg]_{\,q=q_0} = 0\,,
\label{eq:equil-eqs-q0}
\eeq
with the \emph{derived} quantity 
\beq
\mu_0 = \Bigg[\, \frac{\dd \epsilon(q)}{\dd q}\,    \Bigg]_{\,q=q_0}\,.
\label{eq:equil-eqs-mu0}
\eeq
In order for the Minkowski vacuum to be stable, there is the further condition:
$\chi(q_0)>0$ where $\chi$ corresponds to the isothermal vacuum
compressibility \eqref{eq:Stability}~\cite{KlinkhamerVolovik2008a}.
In this equilibrium vacuum the gravitational constant $G(q_0)$ can be identified
with Newton's constant $G_{N}$.

Let us compare the conditions for the equilibrium  self-sustained vacuum, \eqref{eq:equil-eqs-q0} and \eqref{eq:equil-eqs-mu0}, with the two conditions
 suggested by Weinberg, who used the fundamental scalar field $\phi$ for the description of the vacuum.   In this description there are two constant-field equilibrium conditions for Minkowski vacuum, 
$\partial \mathcal{L} / \partial g_{\alpha\beta}=0$
and $\partial \mathcal{L} / \partial \phi =0$, see Eqs.~(6.2) and (6.3) in ~\cite{Weinberg1988}.
These two conditions turn out to be inconsistent,
unless the potential term in $\mathcal{L}(\phi)$ is fine-tuned  (see also Sec.~2 of Ref.~\cite{Weinberg1996}). In other words, the Minkowski vacuum solution may exist only for the  fine-tuned action. This is the Weinberg formulation of the cosmological constant problem.

The self-sustained vacuum naturally bypasses this problem \cite{KlinkhamerVolovik2010}.
Equation~
$\partial \mathcal{L} / \partial g_{\alpha\beta}=0$ corresponds to the   equation~\eqref{eq:equil-eqs-q0}. However, the equation $\partial \mathcal{L} / \partial \phi =0$  is \emph{relaxed} in the $q$-theory of self-sustained vacuum.
Instead of the condition $\partial \mathcal{L} / \partial q =0$,
the conditions are $\nabla_\alpha (\partial \mathcal{L} / \partial q)=0$,
which allow for having $\partial \mathcal{L} / \partial q=\mu$
with an \emph{arbitrary} constant $\mu$.  This is the crucial difference between a fundamental scalar field $\phi$
and the variable $q$ describing the self-sustained vacuum. As a result, the
equilibrium conditions for $g_{\alpha\beta}$ and $q$ can be consistent
without fine-tuning of the original action. For Minkowski vacuum to exist only one condition \eqref{eq:equil-eqs-q0} must be satisfied. In other words, the Minkowski vacuum solution exists for arbitrary action provided that solution of equation \eqref{eq:equil-eqs-q0} exists.

\section{Cosmology as approach to equilibrium} \label{sec:Dynamics}

\subsection{Energy exchange between vacuum and gravity+matter}

In the curved Universe and/or in the presence of matter, $q$ becomes space-time
dependent due to interaction with gravity and matter  (see \eqref{eq:Maxwell2}). As a result the vacuum energy can be  transferred to the energy of gravitational field and/or to the energy of matter fields.
This also means that the energy of matter is not conserved.
The energy-momentum tensor of matter $T^\text{M}_{\mu\nu}$, which enters the generalized Einstein equations \eqref{eq:EinsteinEquationF2},  is
 determined by variation over $g^{\mu\nu}$ at constant $q$. That is why it is not conserved:
\begin{equation}
\nabla_\nu T^{\text{M}\mu\nu} =-\frac{\partial \mathcal{L}^\text{M}}{\partial q}
\nabla_\mu q
 \,.
\label{eq:non-conservation}
\end{equation}
The matter energy can be transferred to the vacuum energy due to interaction
with $q$-field. Using \eqref{eq:Maxwell2} and
the equation \eqref{eq:Lambda} for cosmological constant
one obtains that the vacuum energy is transferred both to gravity
and matter with the rate:
\begin{equation}
\nabla_\mu \Lambda\equiv  \nabla_\mu\rho_{\rm vac}=\left(
\frac{d\epsilon(q)}{d q}-\mu\right)\nabla_\mu q
 =-\frac{R}{16\pi} \frac{dG^{-1}(q)}{d q}   \nabla_\mu q
 +\nabla_\nu T^{\text{M}\mu\nu}
 \,.
\label{eq:non-conservation_vacuum}
\end{equation}

The energy exchange between the vacuum and gravity+matter allows for the relaxation of the vacuum energy and  cosmological ``constant''.

\subsection{Dynamic relaxation of vacuum energy}
\label{Dynamic_relaxation}

Let us assume that 
we can make a sharp kick of the system from its equilibrium state.
For quantum liquids (or any other quantum condensed matter) we know the result of the kick: the liquid or superconductor starts to  relax back to the equilibrium state, and with or without oscillations it finally approaches
the equilibrium
\cite{VolkovKogan1974,Barankov2004,Yuzbashyan2005,Yuzbashyan2008,Gurarie2009}. 
The same should happen with the quantum vacuum.
Let us consider this behavior using  the realization of the vacuum $q$ field in terms of the 4-form field,
when $\mu$ serves as the overall  integration constant. We start with the fully equilibrium vacuum state, which is characterized by the values $q=q_0$ and $\mu=\mu_0$ in \eqref{eq:equil-eqs}.  The kick moves the variable
$q$ away from its equilibrium value, while $\mu$  still remains the same being the overall integration constant, $\mu=\mu_0$. In the non-equilibrium state which arises  immediately after the kick, the vacuum energy is non-zero and big. If the
kick is very sharp, with the time scale of order $t_{\rm P}=1/E_{\rm P}=\sqrt{G_\text{N}}$, the energy density of the vacuum can reach the Planck-scale value,
$\rho_\text{vac}\sim E_\text{P}^4$.

\begin{figure}[t]
\includegraphics[width=.6\textwidth]{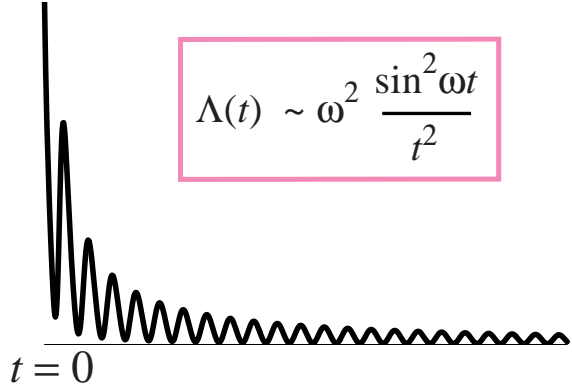}
\caption{
Sketch of the oscillating decay of the cosmological constant after sharp kick. Frequency of oscillations $\omega \sim E_{\rm P}$. If the Universe starts expansion from the state with vacuum energy $\rho_\text{vac}\sim E_\text{P}^4$, in the process of relaxation the vacuum energy will reach the observed value of cosmological constant at present time: $\Lambda(t_{\rm present})\sim  E_\text{P}^2/t_{\rm present}^2\sim \big(10^{-3}\,\text{eV}\big)^4$.
 }
\label{fig:OscillatingDecay}
\end{figure}

For simplicity we ignore  the interaction between the vacuum and matter. Then from the solution
of dynamic equations \eqref{eq:Maxwell2} and \eqref{eq:EinsteinEquationF2} with $\mu=\mu_0$ one 
finds  that after the kick $q$ does  return to  its equilibrium value $q_0$ in  the Minkowski vacuum. At
late time the relaxation has the following asymptotic behavior: \cite{KlinkhamerVolovik2008b} 
\beq\label{eq:time-dependence-q}
q(t)-q_0\sim\,
        q_0  \frac{  \sin\omega\, t }{ \omega\,  t} ~~,~~ \omega \,t\gg 1\,,
\eeq
where oscillation frequency $\omega$ is of the order of the Planck-energy scale
$E_\text{P}$. The gravitational constant $G$ approaches
its Newton value $G_\text{N}$ also with the power-law modulation:
\beq\label{eq:time-dependence-G}
G(t)-G_\text{N} \sim\,
        G_\text{N}    \frac{  \sin\omega\, t }{ \omega\,  t} ~~,~~ \omega \,t\gg 1\,.
\eeq
The vacuum energy relaxes to zero in the following way (see Fig. \ref{fig:OscillatingDecay}):
\bsubeqs\label{eq:VacuumEnergyDecay}
\beqa
 \rho_\text{vac}(t)
 \propto
 \frac{\omega^2}{t^2}\;\sin^2 \omega\,  t~~,~~\omega\,  t\gg 1\,,
\label{eq:VacuumEnergyOscillating-dimensionfull}
\eeqa
For the Planck scale kick,  the vacuum energy density after the kick, i.e.  at $t\sim 1/E_\text{P}$, has a Planck-scale value,
$\rho_\text{vac}\sim E_\text{P}^4$. According to \eqref{eq:VacuumEnergyOscillating-dimensionfull}, 
at present time it must reach the value
\begin{equation}
\overline \rho_\text{vac}(t_{\rm present})
 \propto
 \frac{E_\text{P}^2}{t^2_{\rm present}}\sim E_\text{P}^2H^2 \,,
\end{equation}
where $H$ is the Hubble parameter. This value approximately  corresponds to the
measured value of the cosmological constant.

This, however, can be considered as an illustration of the dynamical reduction of the
large value of the cosmological constant, rather than the real scenario of the evolution of the Universe.
We did not take into account quantum
dissipative effects and the energy exchange between vacuum and matter. 
Indeed, matter field radiation (matter quanta emission)
by the oscillations of the vacuum can be expected to lead to
faster relaxation of the initial vacuum energy~\cite{Starobinsky1980},
\beqa
 \rho_\text{vac}(t)
 \propto \Gamma^4  \exp(-\Gamma\, t)\,,
\label{eq:VacuumEnergyQuantum-dimensionful}
\eeqa
\esubeqs
with a decay rate $\Gamma\sim \omega \sim E_\text{P}$.

Nevertheless, the cancellation mechanism and example of relaxation provide the following lesson. The Minkowski-type solution appears without fine-tuning of the parameters of the action,
precisely because the vacuum is characterized by a constant derivative
of the vacuum field rather than by a constant vacuum field itself.
As a result, the parameter $\mu_0$ emerges in \eqref{eq:equil-eqs-mu}
as an \emph{integration constant}, i.e., as a parameter of the solution
rather than a parameter of the action. Since after the kick the integration constant remains 
intact, the Universe will  return to its equilibrium Minkowski state with $\rho_\text{vac}=0$, even if in the non-equilibrium state after the kick the  vacuum energy could reach
$\rho_\text{vac}\sim E_\text{P}^4$.
The idea that the constant derivative of a field may be important
for the cosmological constant problem has been suggested earlier by
Dolgov~\cite{Dolgov1985,Dolgov1997} and
Polyakov~\cite{Polyakov1991,PolyakovPrivateComm}, where the latter explored
the analogy with the Larkin--Pikin effect~\cite{LarkinPikin1969}
in solid-state physics.

\subsection{Minkowski vacuum as attractor} \label{sec:Attractor}

The example of relaxation of the vacuum energy in Sec. \ref{Dynamic_relaxation} has the principle drawback. 
Instead of the fine-tuning of the action, which is bypassed in the self-sustained vacuum,
we have the fine-tuning  of the integration constant. We assumed that originally the Universe was in its Minkowski ground state,
and thus the specific value of the integration constant $\mu=\mu_0$ has been chosen, that fixes the value $q=q_0$ of the original Minkowski equilibrium vacuum.
In the  4-form realization of the vacuum field, any other choice of the integration constant ($\mu \ne \mu_0$) leads  to a de-Sitter-type solution~\cite{KlinkhamerVolovik2008b}. Though it is not excluded
that the Big-Bang started after the kick from the equilibrium Minkowski vacuum, it is 
instructive to consider the scenarios which avoid this fine-tuning and obtain the natural relaxation
of $\mu$ to $\mu_0$ from any initial state. Such relaxation as we know occurs in quantum liquids. So
we must relax the condition on $\mu$: it should not serve
as an overall integration constant, while remaining the conjugate
variable in thermodynamics. Then using the condensed matter experience one may expect
that the Minkowski equilibrium vacuum becomes an attractor and the  de-Sitter solution with $\mu \neq\mu_0$ will
inevitably relax to Minkowski vacuum with $\mu=\mu_0$.

\begin{figure}[t]
\includegraphics[width=.6\textwidth]{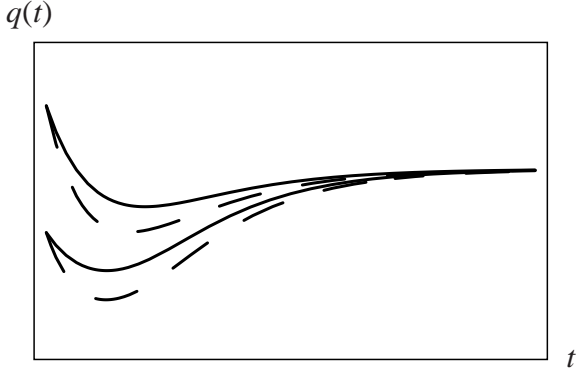}
\caption{
Aaether-field $q$ evolution and Minkowski attractor
 in a spatially flat Friedmann--Robertson--Walker universe in Dolgov model \cite{Dolgov1997} (see \cite{KlinkhamerVolovik2010} for details). 
The  bare cosmological constant is $\Lambda_\text{bare} \sim E_\text{P}^4$.   Four numerical solutions correspond to different boundary conditions, but all approach the Minkowski-spacetime
solution \eqref{eq:asymp_solution}.
The Minkowski vacuum is an attractor
because the vacuum compressibility \eqref{eq:Stability} is positive,
$\chi(q_0)>0$.}
\label{fig:attractor}
\end{figure}
 
 This expectation is confirmed in the aether type realization of the vacuum variable
in terms of a vector field  as discussed in Sec. \ref{Aaether_field}.
The constant vacuum field $q$ there appears 
as the derivative
of a vector field not for all vector fields $u_\beta$, but only for the specific solution $u^q_\beta$
corresponding to the equilibrium vacuum,
$q\,g_{\alpha\beta} \equiv \nabla_\alpha\, u^q_{\beta}=
u^q_{\alpha\beta}$.
In this realization, the effective chemical potential
$\mu \equiv d\epsilon(q)/d q$ appears only for the equilibrium states
(i.e., for their thermodynamical properties),
but $\mu$ does not appear as an integration constant for the dynamics.
Hence, the fine-tuning problem of the integration constant is overcome,
simply because there is no integration constant.

The instability of the de-Sitter solution towards the Minkowski one has been already demonstrated by
Dolgov~\cite{Dolgov1997}, who considered  the simplest
quadratic choices of the Lagrange density of $u_\beta(x)$. 
But his result also holds for the generalized Lagrangian with a generic
function $\epsilon(u_{\alpha\beta})$  in Sec. \ref{Aaether_field} \cite{KlinkhamerVolovik2010}.
In the Dolgov scenario the initial de-Sitter-type expansion evolves towards the Minkowski attractor by the
following $t\rightarrow \infty$ asymptotic solution for the aether-type field
$u_\beta=(u_0(t),0)$ and  Hubble parameter:
\beq
u_0(t)\rightarrow q_0\,t \,,\quad
H(t) \rightarrow 1/t\,.
\label{eq:asymp_solution}
\eeq
At large cosmic times $t$, the curvature terms
decay as $R\sim H^2 \sim 1/t^2$ and the Einstein equations
lead to the nullification of the energy-momentum tensor of the $u_{\beta}$ field:
$T_{\alpha\beta}[u]=0$. Since \eqref{eq:asymp_solution}
with $d u_0/dt = H\,u_0$  satisfies the  $q$--theory \emph{Ansatz}
$u_{\alpha\beta} = q\,g_{\alpha\beta}$,
the  energy-momentum tensor is completely expressed by the single constant $q$:
$T_{\alpha\beta}(q) =[\epsilon(q)  -  q\, d\epsilon(q)/d q]\,g_{\alpha\beta}$.
As a result, the equation  $T_{\alpha\beta}(q)=0$
leads to the equilibrium condition \eqref{eq:equil-eqs-q0}
for the Minkowski vacuum and to the equilibrium value $q=q_0$
in \eqref{eq:asymp_solution}. 

\begin{figure}[t]
\includegraphics[width=.7\textwidth]{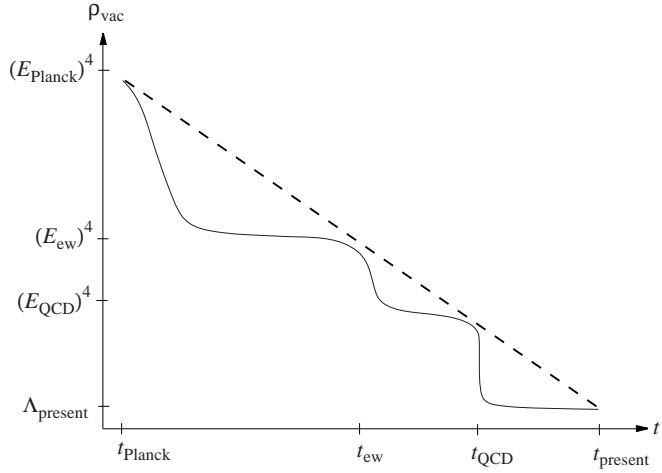}
\caption{
{\it Dashed curve}:  relaxation according to the relation
$<\rho_{\rm vac}(t)>  \;\sim\, (E_{\rm Planck})^2/t^2$ in 
Eq.(\ref{eq:VacuumEnergyOscillating-dimensionfull}).
{\it Full curve}: Sketch of the relaxation of the vacuum energy density
during the evolution of the Universe according to Ref.~\cite{KlinkhamerVolovik2011a}.
 The origin of the current plateau in the vacuum energy $\Lambda_{\rm present}$ 
is discussed in Sec. \ref{sec:Remnant}.
}
\label{fig:StepRelaxation}
\end{figure}

Figure~\ref{fig:attractor} shows explicitly the attractor behavior for the simplest case of Dolgov action,
with the equilibrium value $q_0$ in \eqref{eq:asymp_solution} appearing
\emph{dynamically}.  This simple version of Dolgov scenario does not appear to give a realistic description of the present Universe~\cite{RubakovTinyakov1999} and requires an appropriate modification \cite{EmelyanovKlinkhamer2011}, which demonstrates that the compensation of a
large initial vacuum energy density can occur dynamically
and that Minkowski spacetime can emerge
spontaneously, without setting a chemical potential.
In other words, an ``existence proof'' has been given for the conjecture
that the appropriate Minkowski value $q_0$ can result from an attractor-type
solution of the field equations.
The only condition for the Minkowski vacuum to be an attractor
is a positive vacuum compressibility \eqref{eq:Stability}.

\subsection{Remnant cosmological constant} \label{sec:Remnant}

Figure~\ref{fig:StepRelaxation} demonstrates the possible more
realistic scenario with a
step-wise relaxation of the vacuum energy density \cite{KlinkhamerVolovik2011a}.
The vacuum energy density moves from plateau to plateau responding to the 
possible phase transitions or crossovers in the Standard Model vacuum and
follows, on average, the steadily decreasing matter energy density.
The origin of the current plateau with a small positive value of the
vacuum energy density  $\Lambda_{\rm present}=\rho_{\rm vac}\sim \big(10^{-3}\,\text{eV}\big)^4$ is still not clear. It
may result from the phenomena, which occur in the infrared. It may come for example from anomalies in the neutrino sector of the quantum vacuum, such as non-equilibrium contribution
of the light massive neutrinos to the quantum vacuum \cite{KlinkhamerVolovik2011a}; 
reentrant violation of Lorentz invariance \cite{Volovik2001} and Fermi point splitting in the neutrino sector
\cite{KlinkhamerVolovik2005b,KlinkhamerVolovik2011b}.
The other possible sources include the QCD anomaly 
\cite{Schutzhold2002,KlinkhamerVolovik2009a,UrbanZhitnitsky2009,Ohta2011,Holdom2011};  torsion \cite{Poplawski2011}; relaxation effects during the electroweak crossover \cite{KlinkhamerVolovik2009b}; etc. Most of these scenarios are determined by the momentum space  topology of the quantum vacuum.

\section{Discussion}

To study the problems related to quantum vacuum one must search for the proper extension
of the current theory of elementary particle physics -- the Standard Model -- based on the topology in momentum space.
However, the gravitational properties of the quantum vacuum can be understood even without that: by extending of our experience
with self-sustained macroscopic systems to the quantum vacuum.
A simple picture of quantum vacuum is based on three assumptions:
(i) The quantum vacuum is a self-sustained medium -- the system which is stable at zero external pressure, like quantum liquids.
(ii)  The quantum vacuum is characterized by a conserved charge $q$, which is analog of the particle density $n$ in quantum liquids and which is non-zero in the ground state of the system, $q=q_0\neq 0$.
(iii) The quantum vacuum with $q=q_0$ is with a great precision a Lorentz-invariant state. The latter is the only property which distinguishes the quantum vacuum from the quantum condensed-matter systems (here we consider the properties of the deep vacuum, on the Planck energy scale, and do not discuss subtle effects of spontaneous violation of Lorentz symmetry which in principle may occur in the infrared   
\cite{OPERA2011,KlinkhamerVolovik2005b,KlinkhamerVolovik2011b}).

These assumptions naturally solve the main cosmological constant problem without fine-tuning.
In any self-sustained system, relativistic or non-relativistic, in thermodynamic equilibrium at $T=0$ the zero-point energy of quantum fields is fully compensated by the microscopic degrees of freedom, so that the relevant energy density is zero in the ground state. This consequence of thermodynamics is automatically fulfilled in any system, which may exist  without external environment. This leads to the trivial result for gravity: the cosmological constant in any equilibrium vacuum state is zero.
The zero-point energy  of the Standard Model fields
is automatically compensated by the $q$--field that describes
the degrees of freedom of the deep quantum vacuum.

These assumptions  allow us to suggest that  cosmology is the process of equilibration. From the condensed matter experience we know that the ground state of the system serves as  an attractor:
 starting far away from equilibrium, the quantum liquid finally reaches its ground state. The same should occur in our Universe:  starting far away from equilibrium in a very early phase of universe, the vacuum  is moving towards the Minkowski attractor. We are now close to this attractor, simply because our Universe is old leaving a small remnant cosmological constant measured in present time. 
 The $q$--theory transforms the standard cosmological constant problem into the search for the proper decay mechanism of the vacuum energy density
and for the proper mechanism of formation of small remnant cosmological constant. For that we need the theory of dynamics of quantum vacuum, whose details depend on the topology of the quantum vacuum in momentum space.

\vspace{5mm}

\textbf{Acknowledgements}. This work is supported in part by the Academy of Finland, Centers of
excellence program 2006-2011.

\end{document}